\documentstyle[bo99,epsfig]{article}

\title{Probing Dense Matter in the cores of AGN:
Observations with RXTE and ASCA}
\author{K. A. Weaver}
\affil{NASA/GSFC}

\begin{document}

\maketitle

\begin{abstract}

Preliminary results from an X-ray spectral study of Seyfert 1
galaxies with {\it ASCA} and {\it RXTE} are presented.  
From an analysis of X-ray reprocessing features of Compton reflection
and Fe K$\alpha$ fluorescence, it is found that iron line 
strength is not necessarily a good predictor of the amount 
of reflection. 
The variability properties of Fe K$\alpha$ and reflection do 
not necessarily scale together
and substantial decoupling of the behavior of the reprocessed flux 
with respect to continuum variability is common.  
Such trends suggest the presence of multiple 
and/or complex regions of dense matter
in AGN cores and that standard, simple accretion
disk models drastically oversimplify reality.

\end{abstract}

\section{Introduction}

X-ray observations are excellent for studying the accretion 
process in active galactic nuclei (AGN) 
because they allow us to see directly into the
galaxy core to determine what happens around the black hole,
all the way from the gravitational radius out to scales $\sim$10
billion times larger.
X-rays can arise anywhere there is matter
to reprocess high-energy continuum photons.
Features of X-ray reprocessing such as Fe K$\alpha$ lines and
Compton reflection (Lightman \& White 1988, Guilbert \& Rees 1988),
can arise from an accretion disk (George \& Fabian 1991) or
the obscuring torus of unified models (Krolik, Madau \& Zycki 1994, 
Ghisellini, Haardt \& Matt 1994).  Significant Fe K$\alpha$ emission 
may also be produced in distributions of clouds such 
as the UV/optical broad line region.

Spectral variability is 
critical to understanding where reflection/fluorescence features arise.
If the reprocessing arises in an accretion
disk, we expect it to respond
to the continuum with a lag of only $\sim 3000 M_8$ s, where $M_8$
is the mass of the central black hole in units of $10^8 M_{\odot}$.
If it comes from as far out as an obscuring torus, the lag should
be $\sim 5\times 10^7 L_{44}^{1/2}$ s, where $L_{44}$ is the luminosity in units
of $10^{44}$ erg s$^{-1}$.  For current favored models of Seyfert 1s 
where the observed reprocessing occurs in
a neutral and Compton-thick accretion disk, the iron line and reflection 
hump should be physically coupled. 

\begin{figure}
%
%
%
%
%
\centerline{\psfig{file=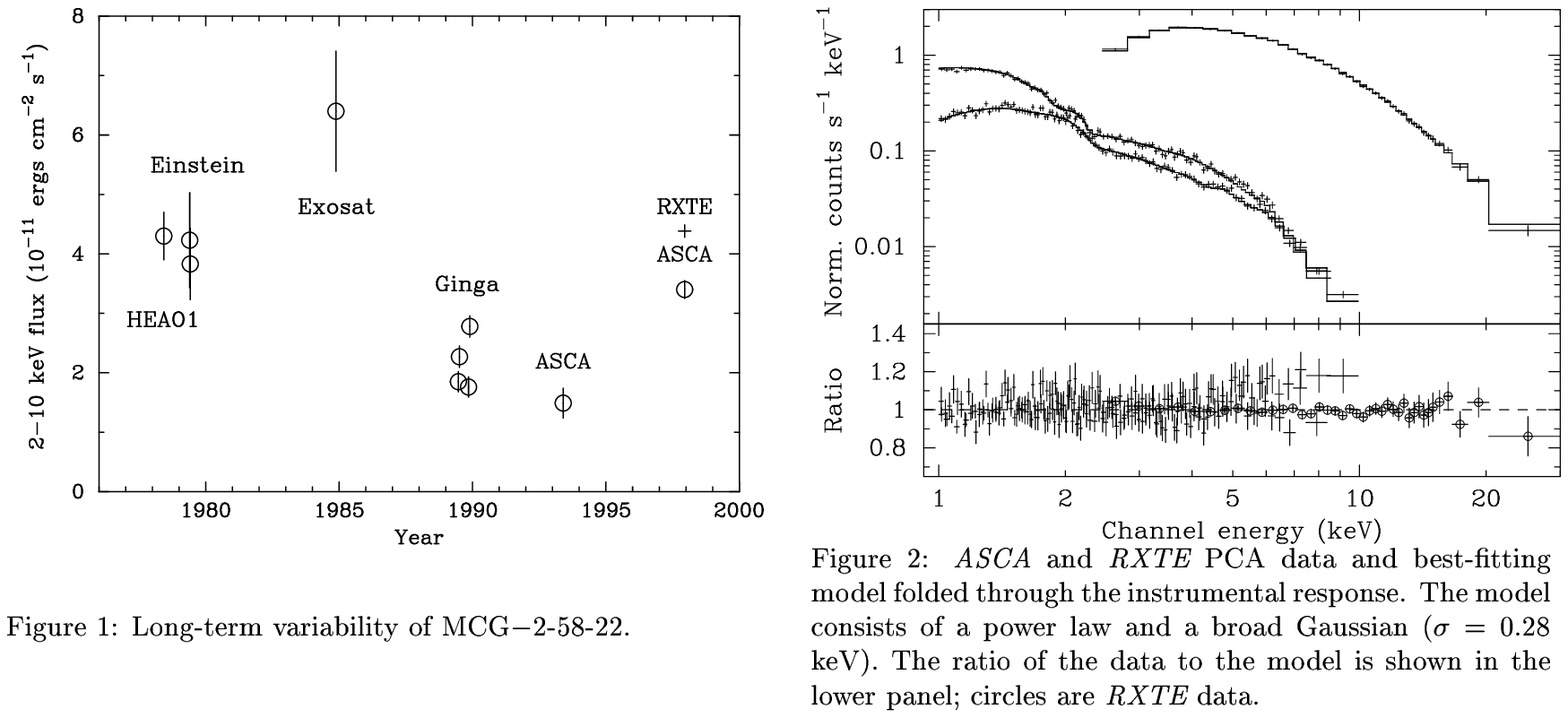, width=12cm,
                   bbllx=0.8in, bblly=7.7in, bburx=7.5in, bbury=10.4in}}
\end{figure}

\subsection{MCG $-$2-58-22: A ``Reflection-less'' Seyfert 1 Galaxy?}

MCG $-$2-58-22 is a bright, unabsorbed Seyfert 1 galaxy with 
evidence for a Doppler broadened iron 
line (FWHM of $31,000^{+30,000}_{-18,000}$ km s$^{-1}$; Weaver et al.\ 1995).
It was observed with {\it ASCA} 
and {\it RXTE} for a total of 38 ks on December 15, 1997.  

MCG $-$2-58-22 varies significantly on timescales of 
years (Figure 1), changing intensity by a factor of three 
between {\it ASCA} observations in 1993 and 1997.  
The {\it ASCA} + {\it RXTE} spectrum and best-fitting
model are shown in Figure 2.  Using the {\it pexrav} model
in {\it xspec} and assuming a face-on disk
(where R = 1 for $\Omega/2\pi$ coverage), {\it RXTE} detects essentially no 
reflection, with an upper limit of R = 0.08.  Comparing {\it ASCA}
observations, the iron line
normalization stays constant at $\sim5\times10^{-5}$ 
photons cm$^{-2}$ s$^{-1}$ and the 
EW drops from $\sim230$ eV to $\sim125$ eV in four years. 

The lack of reflection is puzzling if the broad iron 
line comes from a disk.  Although a large disk inclination or 
overabundance of iron can suppress reflection, the iron 
line would be much broader than observed in the former 
case and much stronger in the latter. 
Relativistic smearing of the 
Compton hump is also not likely because this would smear out 
the line.  Plausible explanations include a partly ionized
accretion disk, a cutoff in the intrinsic spectrum at energies  
between 10 and 40 keV, a Compton-thin accretion disk, 
or a lag in the response 
of reflection to the continuum (larger 
than that of the iron line). 

The spectrum can be more easily explained if the iron line 
originates elsewhere.  The constant line strength suggests  
that the line originates at least a few light years away.
In fact, the line width is marginally consistent with 
the inferred velocity of the UV/optical broad line region 
clouds (16,000 km s$^{-1}$; Wu, Boggess \& Gull 1981).
BLR clouds with $N_{\rm H}\sim10^{23}$ cm$^{-2}$ would   
produce fluorescence but no reflection. 

\begin{figure}
%
%
%
%
%
\centerline{\psfig{file=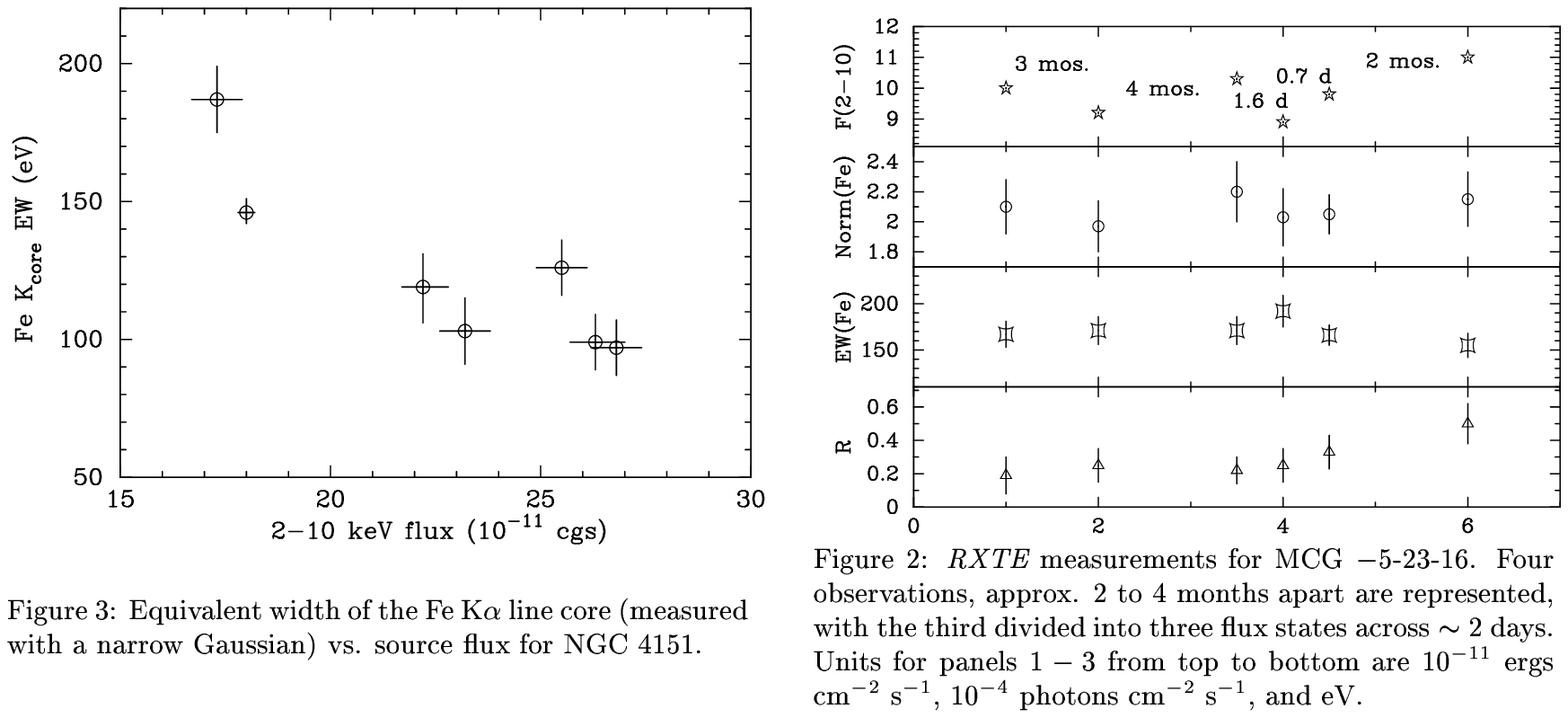, width=12cm,
                   bbllx=0.8in, bblly=7.7in, bburx=7.5in, bbury=10.4in}}
\end{figure}

\subsection{The Contribution of Non-Disk Emission to X-ray Reprocessing}

An obvious complication for spectral studies is the contribution
of reprocessed emission from non-disk regions.
To study this, we have looked at 15 Seyfert 1 galaxies with
multiple observations in the {\it ASCA} archive. 
(Gelbord and Weaver, this volume).  Modeling the iron line 
with a narrow Gaussian (to fit the line core), we find that eleven 
galaxies show 
some evidence for a change in EW.  This can be explained 
as a significant time delay in flux from the line core, which 
might be due to emission from the outer disk,
the BLR or the torus.  One of the 
best examples of non-variability is NGC 4151, for which the 
line core flux does not change and the  
the EW is inversely proportional to the continuum
flux (Figure 3).  

In a second study, we obtained four {\it RXTE} observations of
MCG$-$5-23-16 in 1996 and 1997 spaced $\sim3$ months apart.
The spectral results are plotted in Figure 4. 
Between the third (a long-look covering two days) 
and fourth pointings, the $2-10$ keV flux
decreases while the line normalization
remains the same and the EW drops from about 
200 eV to 150 eV (Mattson and Weaver, in prep).
The lack of response to the continuum suggests that at least some of the line  
emission occurs from far out.  However, the reflection component 
tracks the continuum during the same time.  If 
real, this would strongly support the idea that MCG --5-23-16
has {\it two} physically distinct reprocessing regions. 
Similar behavior of the iron line not tracking 
reflection is seen by {\it RXTE} for
NGC 5548 (Chiang et al.\ 1999).

\begin{table}
  \caption{\protect\small Compton Reflection Fits to RXTE Seyfert 1s.}
  \begin{center}
  \mbox{\protect\small
  \begin{tabular}{ccccc}
  Galaxy      & Date          & PCA rate & R     & F($2-10$)   \\
              &               & (count s$^{-1}$) & & ($10^{-11}$ cgs) \\ \hline
NGC 7469      & 1996 Apr $25-26$  & 9  & 1.10($0.53-1.86$) & 2.9   \\
Mrk 509       & 1996 Nov $04-06$  & 18 & 0.30($0.07-0.45$) & 5.9    \\
MCG $-$5-23-16 & '96 Apr 24 $-$ '97 Jan 10 & 33 & 0.29($0.17-0.42$) & 9.8 \\
IC 4329a      & 1997 Aug 11     & 29 & 0.14($0.00-0.80$) & 10.0 \\
IC 4329a      & 1997 Aug 22     & 47 & 0.46($0.27-0.68$) & 16.0  \\
IC 4329a      & 1997 Sep 01     & 36 & 0.39($0.12-0.78$) & 11.8 \\
IC 4329a      & 1997 Sep 11     & 49 & 0.46($0.22-0.76$) & 17.5 \\
IC 4329a      & 1997 Sep 18     & 30 & 1.04($0.44-1.92$) & 9.9   \\
IC 4329a      & 1997 Oct 02     & 43 & 0.07($0.00-0.45$) & 15.3  \\
MCG $-$2-58-22 & 1997 Dec $15-17$  & 12 & 0.02($0.00-0.08$)* & 3.4   \\ \hline
  \end{tabular} 
}
\end{center}
Errors are 90\% confidence for two parameters. *A 4\% systematic error is included.
\end{table}

\subsection{Compton Reflection and {\it RXTE}}

Preliminary {\it RXTE} and {\it ASCA} results 
suggest that Fe K$\alpha$ emission and 
Compton reflection are at least partly decoupled in their behavior.
Such a result has strong implications and thus
deserves close scrutiny.   In particular, it is important to rule out 
any possible systematic calibration effects.  The PCA detector 
calibration is still being 
refined; however, up to December 1997, the {\it ftools} v4.1 matrix 
for gain epoch 3 (4/15/96 -- 3/22/99) provides 
good fits to the Crab spectrum with less than 4\% residuals
near $6-7$ keV and $10-20$ keV (K. Jahoda, private communication).

Table 1 compares results for a handful of Seyfert 1 galaxies observed
during 1996 and 1997.  Data were fitted with the {\it pexrav} plus 
Gaussian models in {\it xspec} and response matrices were 
generated for each observation with {\it ftools} v4.1. 
MCG --2-58-22 was observed at the end of 1997, where response 
deviations from the Crab begin, but there is no clear 
systematic trend for R to decrease with time.  It is possible 
that R is systematically 
underestimated (the mean for IC 4329a is 0.43, where R=1
for an accretion disk), but this cannot account for the 
non-detection in MCG --2-58-22, even after including a 
4\% systematic error. 
We conclude that there is no time-dependent 
(i.e., calibration-dependent) trend for R 
to decrease with time.   

\section{Conclusions}

Results are summarized from an {\it RXTE} and 
{\it ASCA} investigation of spectral variability in Seyfert 1 galaxies.
The spectra defy a simple interpretation, with significant time 
lags and the Fe K$\alpha$ line 
and Compton reflection behaving in a decoupled fashion. 
A possible explanation is that the spectral variability of the accretion-disk 
is diluted by emission from other, larger regions.  

\begin{acknowledgements}
This work is supported by NASA grants NAG5-7010, NAG5-6917 \&
NAG5-3504.
\end{acknowledgements}

\end{document}